\documentclass[showkeys, showpacs]{revtex4}
\usepackage{indentfirst}
\begin{document}
\title{\bf Molecular state $N\Xi$ in the coupled-channel formalism}
\author{S.M. Gerasyuta}
\email{gerasyuta@SG6488.spb.edu}
\author{E.E. Matskevich}
\email{matskev@pobox.spbu.ru}
\affiliation{Department of Physics, St. Petersburg State Forest Technical
University, Institutski Per. 5, St. Petersburg 194021, Russia}
\begin{abstract}
The relativistic six-quark equations for the molecule $N\Xi$ are found in the
dispersion relation technique. The relativistic six-quark amplitudes of the hexaquark
including the quarks of three flavors ($u$, $d$, $s$) are calculated. The pole
of these amplitudes determines the mass of $N\Xi$ state $M=2252\, MeV$.
The binding energy is equal to $3\, MeV$.
\end{abstract}
\pacs{11.55.Fv, 11.80.Jy, 12.39.Ki, 12.39.Mk.}
\keywords{hexaquarks, dispersion relation technique.}
\maketitle
\section{Introduction.}
The $H$-particle, $N\Omega$-state and di-$\Omega$ may be strong interaction
stable \cite{1}. Jaffe studied the color-magnetic interaction of the
one-gluon-exchange potential in the multiquark system and found the
most attractive channel is the flavor singlet with quark content
$u^2d^2s^2$. The same symmetry analysis of the chiral boson exchange
potential leads to the similar result \cite{2}. Up to now, these three
interesting candidates of dibaryons are still not found or confirmed by
experiments. It seems that one should go beyond these candidates and should
search the possible candidates in a wider region, in terms of a more
reliable model.

There were a number of theoretical predictions by using various models:
the quark-delocalization model \cite{3, 4}, the flavor $SU(3)$ skyrmion
model \cite{5}, the chiral $SU(3)$ quark model \cite{6}, the the quark
cluster model \cite{7, 8}. By employing the chiral $SU(3)$ quark model
Zhang and Yu studied $\Omega\Omega$ and $\Sigma\Omega$ states \cite{9, 10}.
Lomon predicted a deuteron-like dibaryon resonance using R-matrix
theory \cite{11}.

In our previous paper \cite{12} the relativistic six-quark equations are
found in the framework of coupled-channel formalism. The dynamical mixing between
the subamplitudes of hexaquark is considered. The six-quark amplitudes
of dibaryons are calculated. The poles of these amplitudes determine the
masses of dibaryons. We calculated the contribution of six-quark
subamplitudes to the hexaquark amplitudes.

\section{Six-quark amplitudes of molecular state $N\Xi$.}
We derive the relativistic six-quark equations in the framework of the
dispersion relation technique. We use only planar diagrams; the other
diagrams due to the rules of $1/N_c$ expansion \cite{13, 14, 15} are neglected.

We shall consider the derivation of the relativistic generalization of the
Faddeev-Yakubovsky approach \cite{16}. In our case the low-lying dibaryons
are considered. We take into account the pairwise interaction of all six
quarks in the hexaquark.

For instance, we consider the reduced amplitude $\alpha_2^{1^{uu}1^{ss}}$ (Fig. 1).
The system of graphical equations in Fig. 1. determines by the
subamplitudes using the self-consistent method \cite{12}.

The coefficients are determined by the permutation of quarks \cite{17, 18}.
We should use the coefficient multiplying of the diagrams in the graphical
equation Fig. 1.

In Fig. 1 the first coefficient is equal to 4, that the number $4=2$
(permutation of particles 1 and 2) $\times 2$ (permutation of particles
5 and 6); the second coefficient is equal to 4, that the number $4=2$
(permutation of particles 1 and 2) $\times 2$ (permutation of particles
3 and 4); the third coefficient is equal to 4, that the number $4=2$
(permutation of particles 1 and 2) $\times 2$ (permutation of particles
5 and 6); the fourth coefficient is equal to 8, that the number $8=2$
(permutation of particles 1 and 2) $\times 2$ (permutation of particles
3 and 4) $\times 2$ (permutation of particles 5 and 6);
the fifth coefficient is equal to 8, that the number $8=2$
(permutation of particles 1 and 2) $\times 2$ (permutation of particles
3 and 4) $\times 2$ (permutation of particles 5 and 6); the sixth
coefficient is equal to 8, that the number $8=2$
(permutation of particles 1 and 2) $\times 2$ (permutation of particles
3 and 4) $\times 2$ (permutation of particles 5 and 6).

The system of equations for the state $N\Xi$ is given in the Appendix A.

\section{Calculation results.}
The quark masses of model $m_{u,d}=410\, MeV$ and $m_s=557\, MeV$ coincide
with the ordinary baryon ones in our model \cite{19, 20}.
The model in question has only three parameters: the cutoff parameter
$\Lambda=11$ (similar to the three quark model) and the gluon
coupling constants $g_0$ and $g_1$. These parameters are determined by the
$\Lambda\Lambda$ and the di-$\Omega$ masses. We have
considered the two type of calculations \cite{12}. In the first case we use the
gluon coupling constants $g_0=0.653$ (diquark $0^+$) and $g_1=0.292$
(diquark $1^+$), which are fitted by the $\Lambda\Lambda$ state with the
mass $2173\, MeV$ and the di-$\Omega$ with the mass $3232\, MeV$, respectively.
In the second case the gluon coupling constants $g_0=0.647$ and $g_1=0.325$
are determined by the masses of $\Lambda\Lambda$ state with the
$M=2171\, MeV$ and the di-$\Omega$ state $M=3093\, MeV$. The experimental
data of these masses are absent, therefore we use the results of paper \cite{4}.
In our model the correlation of gluon coupling constants $g_0$ and $g_1$ is
similar to the $S$-wave baryon ones. It seems that the first case must prefer
that the deuterium state is described more exact.

\section{Conclusions.}
Jaffe considered the most attractive channel of dibaryons (strangeness $S=-2$,
isospin $I=0$, spin-parity $J^P=0^+$). The molecular state $N\Xi$ ($2252\, MeV$)
with the quantum numbers $SIJ=-2,0,0$ possesses the binding energy $B=3\, MeV$.
The binding energy of deuteron is equal to $B=2.226\pm 0.003\, MeV$.
The calculated subamplitudes $A_i$ are given in the Table \ref{tab1}.
The other interesting states give rise to the following binding energy:
$\Lambda\Lambda$ (the binding energy $B=59\, MeV$) $SIJ=-2,0,0$;
$N\Lambda$ (the binding energy $B=32\, MeV$) $SIJ=-1,\frac{1}{2},1$;
$N\Omega$ (the binding energy $B=39\, MeV$) $SIJ=-3,\frac{1}{2},2$.
The molecule $N\Xi$ will be able to obtain experimentally.

\begin{acknowledgments}
The authors would like to thank T. Barnes for useful discussions.
The work was carried with the support of the Russian Ministry
of Education (grant 2.1.1.68.26) and RFBR, Research Project No. 13-02-91154.
\end{acknowledgments}

\newpage

\appendix

\section{The system of equations for the molecule $N\Xi$.}

\begin{eqnarray}
%1
\alpha_1^{1^{uu}}&=&\lambda+4\, \alpha_1^{0^{ud}} I_1(1^{uu}0^{ud})
+4\, \alpha_1^{0^{us}} I_1(1^{uu}0^{us})
+8\, \alpha_2^{0^{ud}0^{us}} I_2(1^{uu}0^{ud}0^{us})
\, , \\
&&\nonumber\\
%2
\alpha_1^{1^{dd}}&=&\lambda+4\, \alpha_1^{0^{ud}} I_1(1^{dd}0^{ud})
+4\, \alpha_1^{0^{ds}} I_1(1^{dd}0^{ds})
+8\, \alpha_2^{0^{ud}0^{ds}} I_2(1^{dd}0^{ud}0^{ds})
\, , \\
&&\nonumber\\
%3
\alpha_1^{1^{ss}}&=&\lambda+4\, \alpha_1^{0^{us}} I_1(1^{ss}0^{us})
+4\, \alpha_1^{0^{ds}} I_1(1^{ss}0^{ds})
\, , \\
&&\nonumber\\
%4
\alpha_1^{0^{ud}}&=&\lambda+\alpha_1^{1^{uu}} I_1(0^{ud}1^{uu})
+\alpha_1^{1^{dd}} I_1(0^{ud}1^{dd})+2\, \alpha_1^{0^{ud}} I_1(0^{ud}0^{ud})
+2\, \alpha_1^{0^{us}} I_1(0^{ud}0^{us})
+2\, \alpha_1^{0^{ds}} I_1(0^{ud}0^{ds})
\nonumber\\
&&\nonumber\\
&+&2\, \alpha_2^{1^{uu}0^{ds}} I_2(0^{ud}1^{uu}0^{ds})
+2\, \alpha_2^{1^{dd}0^{us}} I_2(0^{ud}1^{dd}0^{us})
+2\, \alpha_2^{0^{ud}0^{us}} I_2(0^{ud}0^{ud}0^{us})
\nonumber\\
&&\nonumber\\
&+&2\, \alpha_2^{0^{ud}0^{ds}} I_2(0^{ud}0^{ud}0^{ds})
\, , \\
&&\nonumber\\
%5
\alpha_1^{0^{us}}&=&\lambda+\alpha_1^{1^{uu}} I_1(0^{us}1^{uu})
+\alpha_1^{1^{ss}} I_1(0^{us}1^{ss})+2\, \alpha_1^{0^{ud}} I_1(0^{us}0^{ud})
+2\, \alpha_1^{0^{us}} I_1(0^{us}0^{us})
+2\, \alpha_1^{0^{ds}} I_1(0^{us}0^{ds})
\nonumber\\
&&\nonumber\\
&+&\alpha_2^{1^{uu}1^{ss}} I_2(0^{us}1^{uu}1^{ss})
+2\, \alpha_2^{1^{uu}0^{ds}} I_2(0^{us}1^{uu}0^{ds})
+2\, \alpha_2^{1^{ss}0^{ud}} I_2(0^{us}1^{ss}0^{ud})
+2\, \alpha_2^{0^{ud}0^{us}} I_2(0^{us}0^{ud}0^{us})
\nonumber\\
&&\nonumber\\
&+&2\, \alpha_2^{0^{ud}0^{ds}} I_2(0^{us}0^{ud}0^{ds})
\, , \\
&&\nonumber\\
%6
\alpha_1^{0^{ds}}&=&\lambda+\alpha_1^{1^{dd}} I_1(0^{ds}1^{dd})
+\alpha_1^{1^{ss}} I_1(0^{ds}1^{ss})+2\, \alpha_1^{0^{ud}} I_1(0^{ds}0^{ud})
+2\, \alpha_1^{0^{us}} I_1(0^{ds}0^{us})
+2\, \alpha_1^{0^{ds}} I_1(0^{ds}0^{ds})
\nonumber\\
&&\nonumber\\
&+&\alpha_2^{1^{dd}1^{ss}} I_2(0^{ds}1^{dd}1^{ss})
+2\, \alpha_2^{1^{dd}0^{us}} I_2(0^{ds}1^{dd}0^{us})
+2\, \alpha_2^{1^{ss}0^{ud}} I_2(0^{ds}1^{ss}0^{ud})
+2\, \alpha_2^{0^{ud}0^{us}} I_2(0^{ds}0^{ud}0^{us})
\nonumber\\
&&\nonumber\\
&+&2\, \alpha_2^{0^{ud}0^{ds}} I_2(0^{ds}0^{ud}0^{ds})
\, , \\
&&\nonumber\\
%7
\label{A7}
\alpha_2^{1^{uu}1^{ss}}&=&\lambda
+4\, \alpha_1^{0^{ud}} I_4(1^{uu}1^{ss}0^{ud})
+4\, \alpha_1^{0^{us}} I_3(1^{uu}1^{ss}0^{us})
+4\, \alpha_1^{0^{ds}} I_4(1^{ss}1^{uu}0^{ds})
\nonumber\\
&&\nonumber\\
&+&8\, \alpha_2^{0^{ud}0^{us}} I_7(1^{ss}1^{uu}0^{us}0^{ud})
+8\, \alpha_2^{0^{ud}0^{ds}} I_6(1^{ss}1^{uu}0^{ds}0^{ud})
+8\, \alpha_3^{0^{ud}0^{us}0^{ds}} I_8(1^{uu}1^{ss}0^{ud}0^{us}0^{ds})
\, , \\
&&\nonumber\\
%8
\alpha_2^{1^{dd}1^{ss}}&=&\lambda
+4\, \alpha_1^{0^{ud}} I_4(1^{dd}1^{ss}0^{ud})
+4\, \alpha_1^{0^{us}} I_4(1^{ss}1^{dd}0^{us})
+4\, \alpha_1^{0^{ds}} I_3(1^{dd}1^{ss}0^{ds})
\nonumber\\
&&\nonumber\\
&+&8\, \alpha_2^{0^{ud}0^{us}} I_6(1^{ss}1^{dd}0^{us}0^{ud})
+8\, \alpha_2^{0^{ud}0^{ds}} I_7(1^{ss}1^{dd}0^{ds}0^{ud})
+8\, \alpha_3^{0^{ud}0^{us}0^{ds}} I_8(1^{dd}1^{ss}0^{ud}0^{ds}0^{us})
\, , \\
&&\nonumber\\
%9
\alpha_2^{1^{uu}0^{ds}}&=&\lambda+\alpha_1^{1^{ss}} I_4(0^{ds}1^{uu}1^{ss})
+2\, \alpha_1^{0^{ud}} (I_3(1^{uu}0^{ds}0^{ud})+I_4(1^{uu}0^{ds}0^{ud}))
+2\, \alpha_1^{0^{us}} I_3(1^{uu}0^{ds}0^{us})
\nonumber\\
&&\nonumber\\
&+&\alpha_2^{1^{dd}1^{ss}} I_5(0^{ds}1^{uu}1^{dd}1^{ss})
+2\, \alpha_2^{1^{ss}0^{ud}} (I_6(1^{uu}0^{ds}0^{ud}1^{ss})
+I_7(1^{uu}0^{ds}0^{ud}1^{ss}))
\nonumber\\
&&\nonumber\\
&+&2\, \alpha_2^{0^{ud}0^{us}} (I_5(1^{uu}0^{ds}0^{ud}0^{us})
+I_7(0^{ds}1^{uu}0^{us}0^{ud}))
+2\, \alpha_2^{0^{ud}0^{ds}} I_6(1^{uu}0^{ds}0^{ud}0^{ds})
\nonumber\\
&&\nonumber\\
&+&2\, \alpha_3^{0^{ud}0^{us}0^{ds}} I_8(1^{uu}0^{ds}0^{ud}0^{us}0^{ds})
\, , \\
&&\nonumber\\
%10
\alpha_2^{1^{dd}0^{us}}&=&\lambda+\alpha_1^{1^{ss}} I_4(0^{us}1^{dd}1^{ss})
+2\, \alpha_1^{0^{ud}} (I_3(1^{dd}0^{us}0^{ud})+I_4(1^{dd}0^{us}0^{ud}))
+2\, \alpha_1^{0^{ds}} I_3(1^{dd}0^{us}0^{ds})
\nonumber\\
&&\nonumber\\
&+&\alpha_2^{1^{uu}1^{ss}} I_5(0^{us}1^{dd}1^{uu}1^{ss})
+2\, \alpha_2^{1^{ss}0^{ud}} (I_6(1^{dd}0^{us}0^{ud}1^{ss})
+I_7(1^{dd}0^{us}0^{ud}1^{ss}))
\nonumber\\
&&\nonumber\\
&+&2\, \alpha_2^{0^{ud}0^{us}} I_6(1^{dd}0^{us}0^{ud}0^{us})
+2\, \alpha_2^{0^{ud}0^{ds}} (I_5(1^{dd}0^{us}0^{ud}0^{ds})
+I_7(0^{us}1^{dd}0^{ds}0^{ud}))
\nonumber\\
&&\nonumber\\
&+&2\, \alpha_3^{0^{ud}0^{us}0^{ds}} I_8(1^{dd}0^{us}0^{ud}0^{ds}0^{us})
\, , \\
&&\nonumber\\
%11
\alpha_2^{1^{ss}0^{ud}}&=&\lambda+\alpha_1^{1^{uu}} I_4(0^{ud}1^{ss}1^{uu})
+\alpha_1^{1^{dd}} I_4(0^{ud}1^{ss}1^{dd})
+2\, \alpha_1^{0^{ud}} I_4(0^{ud}1^{ss}0^{ud})
+2\, \alpha_1^{0^{us}} (I_3(1^{ss}0^{ud}0^{us})
\nonumber\\
&&\nonumber\\
&+&I_4(1^{ss}0^{ud}0^{us}))
+2\, \alpha_1^{0^{ds}} (I_3(1^{ss}0^{ud}0^{ds})+I_4(1^{ss}0^{ud}0^{ds}))
+2\, \alpha_2^{1^{uu}0^{ds}} (I_6(1^{ss}0^{ud}0^{ds}1^{uu})
\nonumber\\
&&\nonumber\\
&+&I_7(1^{ss}0^{ud}0^{ds}1^{uu}))
+2\, \alpha_2^{1^{dd}0^{us}} (I_6(1^{ss}0^{ud}0^{us}1^{dd})
+I_7(1^{ss}0^{ud}0^{us}1^{dd}))
\nonumber\\
&&\nonumber\\
&+&2\, \alpha_2^{0^{ud}0^{us}} (I_6(1^{ss}0^{ud}0^{us}0^{ud})
+I_7(1^{ss}0^{ud}0^{us}0^{ud}))
+2\, \alpha_2^{0^{ud}0^{ds}} (I_6(1^{ss}0^{ud}0^{ds}0^{ud})
\nonumber\\
&&\nonumber\\
&+&I_7(1^{ss}0^{ud}0^{ds}0^{ud}))
+4\, \alpha_3^{0^{ud}0^{us}0^{ds}} I_8(1^{ss}0^{ud}0^{ds}0^{us}0^{ud})
\, , \\
&&\nonumber\\
%12
\alpha_2^{0^{ud}0^{us}}&=&\lambda+\alpha_1^{1^{uu}} I_3(0^{ud}0^{us}1^{uu})
+\alpha_1^{1^{dd}} I_4(0^{ud}0^{us}1^{dd})
+\alpha_1^{1^{ss}} I_4(0^{us}0^{ud}1^{ss})
+\alpha_1^{0^{ud}} (I_3(0^{ud}0^{us}0^{ud})
\nonumber\\
&&\nonumber\\
&+&I_4(0^{ud}0^{us}0^{ud}))
+\alpha_1^{0^{us}} (I_3(0^{ud}0^{us}0^{us})+I_4(0^{us}0^{ud}0^{us}))
+\alpha_1^{0^{ds}} I_3(0^{ud}0^{us}0^{ds})
\nonumber\\
&&\nonumber\\
&+&\alpha_2^{1^{uu}1^{ss}} I_7(0^{ud}0^{us}1^{uu}1^{ss})
+\alpha_2^{1^{dd}1^{ss}} I_6(0^{ud}0^{us}1^{dd}1^{ss})
+\alpha_2^{1^{dd}0^{us}} (I_5(0^{ud}0^{us}0^{us}1^{dd})
\nonumber\\
&&\nonumber\\
&+&I_6(0^{ud}0^{us}1^{dd}0^{us})+I_7(0^{us}0^{ud}0^{us}1^{dd}))
+\alpha_2^{1^{ss}0^{ud}} (I_5(0^{us}0^{ud}0^{ud}1^{ss})
+I_6(0^{ud}0^{us}0^{ud}1^{ss})
\nonumber\\
&&\nonumber\\
&+&I_7(0^{ud}0^{us}0^{ud}1^{ss}))
+\alpha_2^{0^{ud}0^{us}} I_6(0^{ud}0^{us}0^{ud}0^{us})
+\alpha_2^{0^{ud}0^{ds}} (I_5(0^{ud}0^{us}0^{ud}0^{ds})
+I_7(0^{us}0^{ud}0^{ds}0^{ud}))
\nonumber\\
&&\nonumber\\
&+&\alpha_3^{1^{uu}1^{dd}1^{ss}} I_8(0^{ud}0^{us}1^{dd}1^{uu}1^{ss})
+\alpha_3^{0^{ud}0^{us}0^{ds}} I_8(0^{ud}0^{us}0^{ud}0^{ds}0^{us})
\, , \\
&&\nonumber\\
%13
\alpha_2^{0^{ud}0^{ds}}&=&\lambda+\alpha_1^{1^{uu}} I_4(0^{ud}0^{ds}1^{uu})
+\alpha_1^{1^{dd}} I_3(0^{ud}0^{ds}1^{dd})
+\alpha_1^{1^{ss}} I_4(0^{ds}0^{ud}1^{ss})
+\alpha_1^{0^{ud}} (I_3(0^{ud}0^{ds}0^{ud})
\nonumber\\
&&\nonumber\\
&+&I_4(0^{ud}0^{ds}0^{ud}))
+\alpha_1^{0^{us}} I_3(0^{ud}0^{ds}0^{us})
+\alpha_1^{0^{ds}} (I_3(0^{ud}0^{ds}0^{ds})+I_4(0^{ds}0^{ud}0^{ds}))
\nonumber\\
&&\nonumber\\
&+&\alpha_2^{1^{uu}1^{ss}} I_6(0^{ud}0^{ds}1^{uu}1^{ss})
+\alpha_2^{1^{dd}1^{ss}} I_7(0^{ud}0^{ds}1^{dd}1^{ss})
+\alpha_2^{1^{uu}0^{ds}} (I_5(0^{ud}0^{ds}0^{ds}1^{uu})
\nonumber\\
&&\nonumber\\
&+&I_6(0^{ud}0^{ds}1^{uu}0^{ds})+I_7(0^{ds}0^{ud}0^{ds}1^{uu}))
+\alpha_2^{1^{ss}0^{ud}} (I_5(0^{ds}0^{ud}0^{ud}1^{ss})
+I_6(0^{ud}0^{ds}0^{ud}1^{ss})
\nonumber\\
&&\nonumber\\
&+&I_7(0^{ud}0^{ds}0^{ud}1^{ss}))
+\alpha_2^{0^{ud}0^{us}} (I_5(0^{ud}0^{ds}0^{ud}0^{us})
+I_7(0^{ds}0^{ud}0^{us}0^{ud}))
+\alpha_2^{0^{ud}0^{ds}} I_6(0^{ud}0^{ds}0^{ud}0^{ds})
\nonumber\\
&&\nonumber\\
&+&\alpha_3^{1^{uu}1^{dd}1^{ss}} I_8(0^{ud}0^{ds}1^{uu}1^{dd}1^{ss})
+\alpha_3^{0^{ud}0^{us}0^{ds}} I_8(0^{ud}0^{ds}0^{ud}0^{us}0^{ds})
\, , \\
&&\nonumber\\
%14
\alpha_3^{1^{uu}1^{dd}1^{ss}}&=&\lambda
+4\, \alpha_1^{0^{ud}} I_9(1^{uu}1^{dd}1^{ss}0^{ud})
+4\, \alpha_1^{0^{us}} I_9(1^{uu}1^{ss}1^{dd}0^{us})
+4\, \alpha_1^{0^{ds}} I_9(1^{dd}1^{ss}1^{uu}0^{ds})
\nonumber\\
&&\nonumber\\
&+&8\, \alpha_2^{0^{ud}0^{us}} I_{10}(1^{dd}1^{uu}1^{ss}0^{ud}0^{us})
+8\, \alpha_2^{0^{ud}0^{ds}} I_{10}(1^{uu}1^{dd}1^{ss}0^{ud}0^{ds})
\, , \\
&&\nonumber\\
%15
\alpha_3^{0^{ud}0^{us}0^{ds}}&=&\lambda
+\alpha_1^{1^{uu}} I_9(0^{ud}0^{us}0^{ds}1^{uu})
+\alpha_1^{1^{dd}} I_9(0^{ud}0^{ds}0^{us}1^{dd})
+\alpha_1^{1^{ss}} I_9(0^{us}0^{ds}0^{ud}1^{ss})
\nonumber\\
&&\nonumber\\
&+&\alpha_1^{0^{ud}} (I_9(0^{ud}0^{us}0^{ds}0^{ud})
+I_9(0^{ud}0^{ds}0^{us}0^{ud})+I_9(0^{us}0^{ds}0^{ud}0^{ud}))
+\alpha_1^{0^{us}} (I_9(0^{ud}0^{us}0^{ds}0^{us})
\nonumber\\
&&\nonumber\\
&+&I_9(0^{ud}0^{ds}0^{us}0^{us})+I_9(0^{us}0^{ds}0^{ud}0^{us}))
+\alpha_1^{0^{ds}} (I_9(0^{ud}0^{us}0^{ds}0^{ds})
+I_9(0^{ud}0^{ds}0^{us}0^{ds})
\nonumber\\
&&\nonumber\\
&+&I_9(0^{us}0^{ds}0^{ud}0^{ds}))
+\alpha_2^{1^{uu}1^{ss}} I_{10}(0^{ud}0^{us}0^{ds}1^{uu}1^{ss})
+\alpha_2^{1^{dd}1^{ss}} I_{10}(0^{ud}0^{ds}0^{us}1^{dd}1^{ss})
\nonumber\\
&&\nonumber\\
&+&\alpha_2^{1^{uu}0^{ds}} (I_{10}(0^{ud}0^{us}0^{ds}1^{uu}0^{ds})
+I_{10}(0^{us}0^{ud}0^{ds}1^{uu}0^{ds}))
+\alpha_2^{1^{dd}0^{us}} (I_{10}(0^{ud}0^{ds}0^{us}1^{dd}0^{us})
\nonumber\\
&&\nonumber\\
&+&I_{10}(0^{ds}0^{ud}0^{us}1^{dd}0^{us}))
+\alpha_2^{1^{ss}0^{ud}} (I_{10}(0^{ud}0^{us}0^{ds}0^{ud}1^{ss})
+I_{10}(0^{ud}0^{ds}0^{us}0^{ud}1^{ss}))
\nonumber\\
&&\nonumber\\
&+&\alpha_2^{0^{ud}0^{us}} (I_{10}(0^{ud}0^{ds}0^{us}0^{ud}0^{us})
+I_{10}(0^{us}0^{ud}0^{ds}0^{ud}0^{us})
+I_{10}(0^{us}0^{ds}0^{ud}0^{ud}0^{us})
\nonumber\\
&&\nonumber\\
&+&I_{10}(0^{ds}0^{us}0^{ud}0^{ud}0^{us}))
+\alpha_2^{0^{ud}0^{ds}} (I_{10}(0^{ud}0^{us}0^{ds}0^{ud}0^{ds})
+I_{10}(0^{ds}0^{ud}0^{us}0^{ud}0^{ds})
\nonumber\\
&&\nonumber\\
&+&I_{10}(0^{ds}0^{us}0^{ud}0^{ud}0^{ds})
+I_{10}(0^{us}0^{ds}0^{ud}0^{ud}0^{ds}))\, .
\end{eqnarray}

We used the functions $I_1$, $I_2$, $I_3$, $I_4$, $I_5$,
$I_6$, $I_7$, $I_8$, $I_9$, $I_{10}$ \cite{12}:

\begin{eqnarray}
I_1(ij)&=&\frac{B_j(s_0^{13})}{B_i(s_0^{12})}
\int\limits_{(m_1+m_2)^2}^{\frac{(m_1+m_2)^2\Lambda_i}{4}}
\frac{ds'_{12}}{\pi}\frac{G_i^2(s_0^{12})\rho_i(s'_{12})}
{s'_{12}-s_0^{12}} \int\limits_{-1}^{+1} \frac{dz_1(1)}{2}
\frac{1}{1-B_j (s'_{13})}\, , \\
&&\nonumber\\
I_2(ijk)&=&\frac{B_j(s_0^{13}) B_k(s_0^{24})}{B_i(s_0^{12})}
\int\limits_{(m_1+m_2)^2}^{\frac{(m_1+m_2)^2\Lambda_i}{4}}
\frac{ds'_{12}}{\pi}\frac{G_i^2(s_0^{12})\rho_i(s'_{12})}
{s'_{12}-s_0^{12}}
\frac{1}{2\pi}\int\limits_{-1}^{+1}\frac{dz_1(2)}{2}
\int\limits_{-1}^{+1} \frac{dz_2(2)}{2}\nonumber\\
&&\nonumber\\
&\times&
\int\limits_{z_3(2)^-}^{z_3(2)^+} dz_3(2)
\frac{1}{\sqrt{1-z_1^2(2)-z_2^2(2)-z_3^2(2)+2z_1(2) z_2(2) z_3(2)}}
\nonumber\\
&&\nonumber\\
&\times& \frac{1}{1-B_j (s'_{13})} \frac{1}{1-B_k (s'_{24})}
 \, , \\
&&\nonumber\\
I_3(ijk)&=&\frac{B_k(s_0^{23})}{B_i(s_0^{12}) B_j(s_0^{34})}
\int\limits_{(m_1+m_2)^2}^{\frac{(m_1+m_2)^2\Lambda_i}{4}}
\frac{ds'_{12}}{\pi}\frac{G_i^2(s_0^{12})\rho_i(s'_{12})}
{s'_{12}-s_0^{12}}\nonumber\\
&&\nonumber\\
&\times&\int\limits_{(m_3+m_4)^2}^{\frac{(m_3+m_4)^2\Lambda_j}{4}}
\frac{ds'_{34}}{\pi}\frac{G_j^2(s_0^{34})\rho_j(s'_{34})}
{s'_{34}-s_0^{34}}
\int\limits_{-1}^{+1} \frac{dz_1(3)}{2} \int\limits_{-1}^{+1}
\frac{dz_2(3)}{2} \frac{1}{1-B_k (s'_{23})} \, , \\
&&\nonumber\\
I_4(ijk)&=&I_1(ik) \, , \\
&&\nonumber\\
I_5(ijkl)&=&I_2(ikl) \, , \\
&&\nonumber\\
I_6(ijkl)&=&I_1(ik) \cdot I_1(jl)
 \, , \\
&&\nonumber\\
I_7(ijkl)&=&\frac{B_k(s_0^{23})B_l(s_0^{45})}{B_i(s_0^{12}) B_j(s_0^{34})}
\int\limits_{(m_1+m_2)^2}^{\frac{(m_1+m_2)^2\Lambda_i}{4}}
\frac{ds'_{12}}{\pi}\frac{G_i^2(s_0^{12})\rho_i(s'_{12})}
{s'_{12}-s_0^{12}}\nonumber\\
&&\nonumber\\
&\times&\int\limits_{(m_3+m_4)^2}^{\frac{(m_3+m_4)^2\Lambda_j}{4}}
\frac{ds'_{34}}{\pi}\frac{G_j^2(s_0^{34})\rho_j(s'_{34})}
{s'_{34}-s_{34}}
\frac{1}{2\pi}\int\limits_{-1}^{+1}\frac{dz_1(7)}{2}
\int\limits_{-1}^{+1} \frac{dz_2(7)}{2}
\int\limits_{-1}^{+1} \frac{dz_3(7)}{2}
\nonumber\\
&&\nonumber\\
&\times&
\int\limits_{z_4(7)^-}^{z_4(7)^+} dz_4(7)
\frac{1}{\sqrt{1-z_1^2(7)-z_3^2(7)-z_4^2(7)+2z_1(7) z_3(7) z_4(7)}}
\nonumber\\
&&\nonumber\\
&\times& \frac{1}{1-B_k (s'_{23})} \frac{1}{1-B_l (s'_{45})}
 \, , \\
&&\nonumber\\
I_8(ijklm)&=&\frac{B_k(s_0^{15})B_l(s_0^{23})B_m(s_0^{46})}
{B_i(s_0^{12}) B_j(s_0^{34})}
\int\limits_{(m_1+m_2)^2}^{\frac{(m_1+m_2)^2\Lambda_i}{4}}
\frac{ds'_{12}}{\pi}\frac{G_i^2(s_0^{12})\rho_i(s'_{12})}
{s'_{12}-s_0^{12}}\nonumber\\
&&\nonumber\\
&\times&\int\limits_{(m_3+m_4)^2}^{\frac{(m_3+m_4)^2\Lambda_j}{4}}
\frac{ds'_{34}}{\pi}\frac{G_j^2(s_0^{34})\rho_j(s'_{34})}
{s'_{34}-s_0^{34}}\nonumber\\
&&\nonumber\\
&\times&\frac{1}{(2\pi)^2}\int\limits_{-1}^{+1}\frac{dz_1(8)}{2}
\int\limits_{-1}^{+1} \frac{dz_2(8)}{2}
\int\limits_{-1}^{+1} \frac{dz_3(8)}{2}
\int\limits_{z_4(8)^-}^{z_4(8)^+} dz_4(8)
\int\limits_{-1}^{+1} \frac{dz_5(8)}{2}
\int\limits_{z_6(8)^-}^{z_6(8)^+} dz_6(8)
\nonumber\\
&&\nonumber\\
&\times&
\frac{1}{\sqrt{1-z_1^2(8)-z_3^2(8)-z_4^2(8)+2z_1(8) z_3(8) z_4(8)}}
\nonumber\\
&&\nonumber\\
&\times&
\frac{1}{\sqrt{1-z_2^2(8)-z_5^2(8)-z_6^2(8)+2z_2(8) z_5(8) z_6(8)}}
\nonumber\\
&&\nonumber\\
&\times& \frac{1}{1-B_k (s'_{15})} \frac{1}{1-B_l (s'_{23})}
\frac{1}{1-B_m (s'_{46})}
 \, , \\
&&\nonumber\\
I_9(ijkl)&=&I_3(ijl) \, , \\
&&\nonumber\\
I_{10}(ijklm)&=
&\frac{B_l(s_0^{23})B_m(s_0^{45})}
{B_i(s_0^{12}) B_j(s_0^{34}) B_k(s_0^{56})}
\int\limits_{(m_1+m_2)^2}^{\frac{(m_1+m_2)^2\Lambda_i}{4}}
\frac{ds'_{12}}{\pi}\frac{G_i^2(s_0^{12})\rho_i(s'_{12})}{s'_{12}-s_0^{12}}
\nonumber\\
&&\nonumber\\
&\times&
\int\limits_{(m_3+m_4)^2}^{\frac{(m_3+m_4)^2\Lambda_j}{4}}
\frac{ds'_{34}}{\pi}\frac{G_j^2(s_0^{34})\rho_j(s'_{34})}
{s'_{34}-s_0^{34}}
\int\limits_{(m_5+m_6)^2}^{\frac{(m_5+m_6)^2\Lambda_k}{4}}
\frac{ds'_{56}}{\pi}\frac{G_k^2(s_0^{56})\rho_k(s'_{56})}{s'_{56}-s_0^{56}}
\nonumber\\
&&\nonumber\\
&\times&
\frac{1}{2\pi}\int\limits_{-1}^{+1}\frac{dz_1(10)}{2}
\int\limits_{-1}^{+1} \frac{dz_2(10)}{2}
\int\limits_{-1}^{+1} \frac{dz_3(10)}{2}
\int\limits_{-1}^{+1} \frac{dz_4(10)}{2}
\int\limits_{z_5(1-)^-}^{z_5(10)^+} dz_5(10)
\nonumber\\
&&\nonumber\\
&\times&
\frac{1}{\sqrt{1-z_1^2(10)-z_4^2(10)-z_5^2(10)+2z_1(10) z_4(10) z_5(10)}}
\nonumber\\
&&\nonumber\\
&\times& \frac{1}{1-B_l (s'_{23})} \frac{1}{1-B_m (s'_{45})}
 \, .
\end{eqnarray}

\newpage

\begin{picture}(600,90)
%1
\put(0,45){\line(1,0){18}}
\put(0,47){\line(1,0){17.5}}
\put(0,49){\line(1,0){17}}
\put(0,51){\line(1,0){17}}
\put(0,53){\line(1,0){17.5}}
\put(0,55){\line(1,0){18}}
\put(30,50){\circle{25}}
\put(19,46){\line(1,1){15}}
\put(22,41){\line(1,1){17}}
\put(27.5,38.5){\line(1,1){14}}
\put(31,63){\vector(1,1){20}}
\put(31,38){\vector(1,-1){20}}
\put(47.5,60){\circle{16}}
\put(47.5,40){\circle{16}}
\put(55,64){\vector(3,2){18}}
\put(55,36){\vector(3,-2){18}}
\put(55,64){\vector(3,-2){18}}
\put(55,36){\vector(3,2){18}}
\put(78,75){1}
\put(64,77){$u$}
\put(78,53){2}
\put(69,58){$u$}
\put(78,41){3}
\put(69,38){$s$}
\put(78,18){4}
\put(65,16){$s$}
\put(54,80){5}
\put(37,80){$d$}
\put(54,13){6}
\put(37,12){$d$}
\put(41.5,56){\small $1^{uu}$}
\put(41.5,36){\small $1^{ss}$}
\put(90,48){$=$}
\put(30,-15){$\alpha_2^{1^{uu}1^{ss}}$}
%2
\put(110,45){\line(1,0){19}}
\put(110,47){\line(1,0){21}}
\put(110,49){\line(1,0){23}}
\put(110,51){\line(1,0){23}}
\put(110,53){\line(1,0){21}}
\put(110,55){\line(1,0){19}}
\put(140,60){\circle{16}}
\put(140,40){\circle{16}}
\put(147.5,64){\vector(3,2){18}}
\put(147.5,36){\vector(3,-2){18}}
\put(147.5,64){\vector(3,-2){18}}
\put(147.5,36){\vector(3,2){18}}
\put(128,55){\vector(1,3){11}}
\put(128,45){\vector(1,-3){11}}
\put(170,75){1}
\put(156,77){$u$}
\put(170,53){2}
\put(161,58){$u$}
\put(170,41){3}
\put(161,38){$s$}
\put(170,18){4}
\put(157,16){$s$}
\put(143,86){5}
\put(126,84){$d$}
\put(143,08){6}
\put(126,9){$d$}
\put(134,57){\small $1^{uu}$}
\put(134,37){\small $1^{ss}$}
\put(140,-15){$\lambda$}
%3
\put(183,48){$+$}
\put(199,48){4}
\put(212,45){\line(1,0){18}}
\put(212,47){\line(1,0){17.5}}
\put(212,49){\line(1,0){17}}
\put(212,51){\line(1,0){17}}
\put(212,53){\line(1,0){17.5}}
\put(212,55){\line(1,0){18}}
\put(242,50){\circle{25}}
\put(231,46){\line(1,1){15}}
\put(234,41){\line(1,1){17}}
\put(239.5,38.5){\line(1,1){14}}
\put(258,62){\circle{16}}
\put(263.5,68.5){\vector(1,1){15}}
\put(263.5,68.5){\vector(1,-1){18}}
\put(254,50){\vector(1,0){28}}
\put(290,50){\circle{16}}
\put(298,50){\vector(3,1){22}}
\put(298,50){\vector(3,-1){22}}
\put(277,59){1}
\put(270,65){$u$}
\put(271,39){2}
\put(265,52){$u$}
\put(314,62){1}
\put(304,58){$u$}
\put(314,30){2}
\put(304,36){$u$}
\put(280,85){5}
\put(267,84){$d$}
\put(258,37){\circle{16}}
\put(265,32){\vector(3,-1){20}}
\put(265,32){\vector(2,-3){12}}
\put(243,38){\vector(1,-3){8}}
\put(288,22){3}
\put(280,31){$s$}
\put(279,7){4}
\put(268,9){$s$}
\put(253,7){6}
\put(240,9){$d$}
\put(252,59){\small $0^{ud}$}
\put(284,47){\small $1^{uu}$}
\put(252,34){\small $1^{ss}$}
\put(230,-15){$4\, \alpha_1^{0^{ud}}\, I_4(1^{uu}1^{ss}0^{ud})$}
%4
\put(335,48){$+$}
\put(353,48){4}
\put(368,45){\line(1,0){18}}
\put(368,47){\line(1,0){17.5}}
\put(368,49){\line(1,0){17}}
\put(368,51){\line(1,0){17}}
\put(368,53){\line(1,0){17.5}}
\put(368,55){\line(1,0){18}}
\put(398,50){\circle{25}}
\put(387,46){\line(1,1){15}}
\put(390,41){\line(1,1){17}}
\put(395.5,38.5){\line(1,1){14}}
\put(399,63){\vector(1,1){20}}
\put(399,38){\vector(1,-1){20}}
\put(422,80){5}
\put(406,80){$d$}
\put(422,13){6}
\put(406,13){$d$}
\put(418,50){\circle{16}}
\put(426,50){\vector(3,2){17}}
\put(426,50){\vector(3,-2){17}}
\put(403,61){\vector(1,0){40}}
\put(403,39){\vector(1,0){40}}
\put(451,60){\circle{16}}
\put(451,40){\circle{16}}
\put(459,61){\vector(3,1){20}}
\put(459,61){\vector(3,-1){20}}
\put(459,39){\vector(3,1){20}}
\put(459,39){\vector(3,-1){20}}
\put(476,72){1}
\put(468,70){$u$}
\put(481,56){2}
\put(461,52){$u$}
\put(481,40){3}
\put(461,44){$s$}
\put(476,22){4}
\put(468,24){$s$}
\put(426,66){1}
\put(414,66){$u$}
\put(426,54){\small 2}
\put(437,51){$u$}
\put(426,40){\small 3}
\put(437,44){$s$}
\put(426,28){4}
\put(415,29){$s$}
\put(412,47){\small $0^{us}$}
\put(445,57){\small $1^{uu}$}
\put(445,37){\small $1^{ss}$}
\put(380,-15){$4\, \alpha_1^{0^{us}}\, I_3(1^{uu}1^{ss}0^{us})$}
\end{picture}

\vskip60pt
\begin{picture}(600,90)
%1
\put(90,48){$+$}
\put(107,48){4}
\put(125,45){\line(1,0){18}}
\put(125,47){\line(1,0){17.5}}
\put(125,49){\line(1,0){17}}
\put(125,51){\line(1,0){17}}
\put(125,53){\line(1,0){17.5}}
\put(125,55){\line(1,0){18}}
\put(155,50){\circle{25}}
\put(144,46){\line(1,1){15}}
\put(147,41){\line(1,1){17}}
\put(152.5,38.5){\line(1,1){14}}
\put(171,62){\circle{16}}
\put(176.5,68.5){\vector(1,1){15}}
\put(176.5,68.5){\vector(1,-1){18}}
\put(167,50){\vector(1,0){28}}
\put(203,50){\circle{16}}
\put(211,50){\vector(3,1){22}}
\put(211,50){\vector(3,-1){22}}
\put(190,59){1}
\put(183,65){$s$}
\put(184,39){2}
\put(180,52){$s$}
\put(227,62){1}
\put(218,59){$s$}
\put(227,30){2}
\put(218,35){$s$}
\put(193,85){5}
\put(180,84){$d$}
\put(171,37){\circle{16}}
\put(178,32){\vector(3,-1){20}}
\put(178,32){\vector(2,-3){12}}
\put(156,38){\vector(1,-3){8}}
\put(201,22){3}
\put(192,31){$u$}
\put(192,7){4}
\put(179,9){$u$}
\put(166,7){6}
\put(153,8){$d$}
\put(165,59){\small $0^{ds}$}
\put(197,47){\small $1^{ss}$}
\put(165,34){\small $1^{uu}$}
\put(137,-15){$4\, \alpha_1^{0^{ds}}\, I_4(1^{ss}1^{uu}0^{ds})$}
%2
\put(250,48){$+$}
\put(267,48){8}
\put(285,45){\line(1,0){18}}
\put(285,47){\line(1,0){17.5}}
\put(285,49){\line(1,0){17}}
\put(285,51){\line(1,0){17}}
\put(285,53){\line(1,0){17.5}}
\put(285,55){\line(1,0){18}}
\put(315,50){\circle{25}}
\put(304,46){\line(1,1){15}}
\put(307,41){\line(1,1){17}}
\put(312.5,38.5){\line(1,1){14}}
\put(334,57){\circle{16}}
\put(331,37){\circle{16}}
\put(341,61){\vector(1,1){15}}
\put(316,62.5){\vector(3,1){40}}
\put(364,77){\circle{16}}
\put(341,61){\vector(1,-1){12}}
\put(339,35){\vector(1,1){14}}
\put(361,50){\circle{16}}
\put(369,50){\vector(3,2){18}}
\put(369,50){\vector(3,-2){18}}
\put(372,78){\vector(3,2){18}}
\put(372,78){\vector(3,-2){18}}
\put(339,35){\vector(1,-1){16}}
\put(316,38){\vector(1,-2){11}}
\put(395,87){1}
\put(376,88){$s$}
\put(395,64){2}
\put(382,74){$s$}
\put(391,54){3}
\put(380,50){$u$}
\put(391,32){4}
\put(376,33){$u$}
\put(345,80){1}
\put(330,75){$s$}
\put(352,63){2}
\put(340,65){$s$}
\put(342,46){3}
\put(348,57){$u$}
\put(345,32){4}
\put(350,39){$u$}
\put(358,11){5}
\put(344,12){$d$}
\put(330,9){6}
\put(316,10){$d$}
\put(328,54){\small $0^{us}$}
\put(325,34){\small $0^{ud}$}
\put(358,74){\small $1^{ss}$}
\put(355,47){\small $1^{uu}$}
\put(290,-15){$8\, \alpha_2^{0^{ud}0^{us}}\, I_7(1^{ss}1^{uu}0^{us}0^{ud})$}
\end{picture}

\vskip60pt
\begin{picture}(600,90)
%1
\put(90,48){$+$}
\put(107,48){8}
\put(125,45){\line(1,0){18}}
\put(125,47){\line(1,0){17.5}}
\put(125,49){\line(1,0){17}}
\put(125,51){\line(1,0){17}}
\put(125,53){\line(1,0){17.5}}
\put(125,55){\line(1,0){18}}
\put(155,50){\circle{25}}
\put(144,46){\line(1,1){15}}
\put(147,41){\line(1,1){17}}
\put(152.5,38.5){\line(1,1){14}}
\put(164,68){\circle{16}}
\put(164,32){\circle{16}}
\put(167,53){\vector(3,1){28}}
\put(167,47){\vector(3,-1){28}}
\put(172,70){\vector(3,2){21}}
\put(172,70){\vector(3,-1){23}}
\put(172,30){\vector(3,1){23}}
\put(172,30){\vector(3,-2){21}}
\put(203,60){\circle{16}}
\put(203,40){\circle{16}}
\put(212,60){\vector(3,2){21}}
\put(212,60){\vector(3,-1){23}}
\put(212,40){\vector(3,1){23}}
\put(212,40){\vector(3,-2){21}}
\put(239,71){1}
\put(222,76){$s$}
\put(240,52){2}
\put(226,59){$s$}
\put(240,42){3}
\put(226,38){$u$}
\put(239,19){4}
\put(222,21){$u$}
\put(183,86){5}
\put(173,82){$d$}
\put(183,7){6}
\put(173,9){$d$}
\put(188,68){1}
\put(181,69){$s$}
\put(188,51){2}
\put(176,59){$s$}
\put(188,42){3}
\put(176,36){$u$}
\put(188,25){4}
\put(181,26){$u$}
\put(158,65){\small $0^{ds}$}
\put(158,29){\small $0^{ud}$}
\put(197,57){\small $1^{ss}$}
\put(197,37){\small $1^{uu}$}
\put(130,-15){$8\, \alpha_2^{0^{ud}0^{ds}}\, I_6(1^{ss}1^{uu}0^{ds}0^{ud})$}
%2
\put(252,48){$+$}
\put(269,48){8}
\put(285,45){\line(1,0){18}}
\put(285,47){\line(1,0){17.5}}
\put(285,49){\line(1,0){17}}
\put(285,51){\line(1,0){17}}
\put(285,53){\line(1,0){17.5}}
\put(285,55){\line(1,0){18}}
\put(315,50){\circle{25}}
\put(304,46){\line(1,1){15}}
\put(307,41){\line(1,1){17}}
\put(312.5,38.5){\line(1,1){14}}
\put(324,68){\circle{16}}
\put(324,32){\circle{16}}
\put(335,50){\circle{16}}
\put(343,50){\vector(1,1){12}}
\put(343,50){\vector(1,-1){12}}
\put(332,70){\vector(3,2){21}}
\put(332,70){\vector(3,-1){23}}
\put(332,30){\vector(3,1){23}}
\put(332,30){\vector(3,-2){21}}
\put(363,60){\circle{16}}
\put(363,40){\circle{16}}
\put(372,60){\vector(3,2){21}}
\put(372,60){\vector(3,-1){23}}
\put(372,40){\vector(3,1){23}}
\put(372,40){\vector(3,-2){21}}
\put(399,71){1}
\put(380,74){$u$}
\put(400,52){2}
\put(387,58){$u$}
\put(400,42){3}
\put(387,37){$s$}
\put(399,19){4}
\put(380,23){$s$}
\put(343,86){5}
\put(333,82){$d$}
\put(343,7){6}
\put(333,11){$d$}
\put(348,68){1}
\put(341,69){$u$}
\put(342,56){2}
\put(350,51){$u$}
\put(342,37){3}
\put(351,45){$s$}
\put(348,25){4}
\put(341,26){$s$}
\put(318,65){\small $0^{ud}$}
\put(318,29){\small $0^{ds}$}
\put(329,47){\small $0^{us}$}
\put(357,57){\small $1^{uu}$}
\put(357,37){\small $1^{ss}$}
\put(275,-15){$8\, \alpha_3^{0^{ud}0^{us}0^{ds}}\, I_8(1^{uu}1^{ss}0^{ud}0^{us}0^{ds})$}
\put(20,-50){Fig. 1. The graphical equations of the reduced amplitude $\alpha_2^{1^{uu}1^{ss}}$ (\ref{A7}).}
\end{picture}

\newpage

\begin{table}
\caption{$N\Xi$ ($2252\, MeV$) $(SIJ=-200)$.
Parameters of model: cutoff $\Lambda=11.0$,
gluon coupling constants $g_0=0.653$ and $g_1=0.292$. Quark masses
$m_{u, d}=410\, MeV$ and $m_s=557\, MeV$.}
\label{tab1}
\begin{tabular}{|c|c|}
\hline
Subamplitudes & Contributions, percent \\
\hline
$A_1^{1^{uu}}$ & 1.96 \\
$A_1^{1^{dd}}$ & 1.96 \\
$A_1^{1^{ss}}$ & 2.56 \\
$A_1^{0^{ud}}$ & 4.33 \\
$A_1^{0^{us}}$ & 5.30 \\
$A_1^{0^{ds}}$ & 5.30 \\
$A_2^{1^{uu}1^{ss}}$ & 7.59 \\
$A_2^{1^{dd}1^{ss}}$ & 7.59 \\
$A_2^{1^{uu}0^{ds}}$ & 4.93 \\
$A_2^{1^{dd}0^{us}}$ & 4.93 \\
$A_2^{1^{ss}0^{ud}}$ & 12.95 \\
$A_2^{0^{ud}0^{us}}$ & 11.87 \\
$A_2^{0^{ud}0^{ds}}$ & 11.87 \\
$A_3^{1^{uu}1^{dd}1^{ss}}$ & 4.10 \\
$A_3^{0^{ud}0^{us}0^{ds}}$ & 12.75 \\
\hline
$\sum A_1$ & 21.41 \\
$\sum A_2$ & 61.73 \\
$\sum A_3$ & 16.86 \\
\hline
\end{tabular}
\end{table}

\end{document}